\documentclass{article}

\usepackage{natbib}
\setcitestyle{numbers,square}

    \usepackage[final]{neurips_2023}

\usepackage{amsmath,amsfonts,amssymb,amsthm,mathrsfs,bm} 
\usepackage[utf8]{inputenc} 
\usepackage[T1]{fontenc}    
\usepackage{hyperref}       
\usepackage{url}            
\usepackage{booktabs}       
\usepackage{nicefrac}       
\usepackage{microtype}      
\usepackage{xcolor}         
\usepackage{graphicx}
\usepackage{multirow}
\usepackage{pifont}
\newcommand{\xmark}{\ding{55}}%

\usepackage[capitalize]{cleveref}
\crefname{section}{Sec.}{Secs.}
\Crefname{section}{Section}{Sections}
\Crefname{table}{Table}{Tables}
\crefname{table}{Tab.}{Tabs.}
\crefname{figure}{Fig.}{Figs.}

\title{Simplicial Message Passing for 
Chemical Property Prediction
}

\author{%
  Hai Lan \\
  Fujian Institute of Research on the Structure of Matter\\
  Chinese Academy of Sciences\\
  \texttt{lanhai09@mails.ucas.ac.cn} \\
  \And
  Xian Wei\thanks{Corresponding Author} \\
  Software Engineering Department\\
  East China Normal University \\
  \texttt{xian.wei@tum.de} \\
}

\newtheorem{definition}{Definition}
\DeclareUnicodeCharacter{2212}{-}

\begin{document}

\maketitle

\begin{abstract}
Recently, message-passing Neural networks (MPNN)
provide a promising tool for dealing with molecular graphs and have achieved remarkable success in facilitating the discovery and materials design with desired properties.
%
%
%
However, the classical MPNN methods also suffer from a limitation in capturing the strong topological information hidden in molecular structures, such as nonisomorphic graphs.
%
To address this problem, this work proposes 
a Simplicial Message Passing (SMP) framework
to better capture the topological information from molecules, which can break through the limitation within the vanilla message-passing paradigm. 
In SMP, a generalized message-passing framework is established for aggregating the information from arbitrary-order simplicial complex, and a hierarchical structure is elaborated to allow information exchange between different order simplices. 
We apply the SMP framework within deep learning architectures for quantum-chemical properties prediction and achieve state-of-the-art results. The results show that compared to traditional MPNN, involving higher-order simplex 
can better capture the complex structure of molecules and
substantially enhance the performance of tasks. 
The SMP-based model can provide a generalized framework for GNNs and aid in the discovery and design of materials with tailored properties for various applications.
\end{abstract}

\section{Introduction}
Capturing topological information is crucial for predicting the properties of functional materials\cite{overview2}. The topological structure of materials, such as molecules or crystals, contains valuable information about their relationships, bond patterns, and atomic arrangements\cite{atz2021geometric,ryan2018crystal}. For example, the topological arrangement of atoms in a crystal or molecule affects its electronic structure and energetics, which plays a vital role in understanding how materials interact with each other and undergo chemical reactions. 
Topological information determines the physical and chemical properties of materials. 
By capturing the topological information, the complex system can be represented in a way that preserves the essential geometric and connectivity features\cite{kahle2014topology}, which can more efficiently establish property-structure relationships, and facilitate the prediction and design of materials with desired properties.

\begin{figure}[h]
  \centering
  \includegraphics[width=\linewidth]{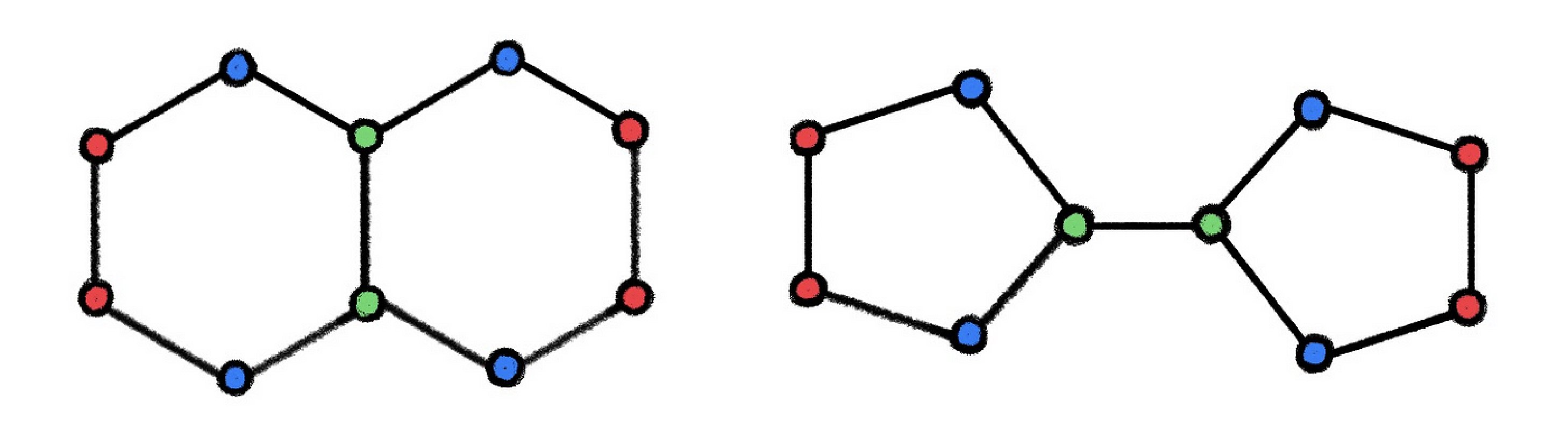}
  \caption{WL graph isomorphism test fails to distinguish two non-isomorphic graphs.}
  \label{fig1}
\end{figure}

To capture the topological information, one common approach is to represent molecules as graphs, where atoms are nodes and chemical bonds are edges\cite{molecularfingerprints2015,schnet20171, MPNN2017,spookynet2021}. 
Each node can be associated with atom-specific features such as atomic number, hybridization state, or atomic mass. Edge features often include the type of bond, the length of the bond, or the angle of the bond. 
Compared to molecular fingerprints, such as Extended Connectivity Fingerprints (ECFP) and Simplified Molecular Input Line Entry System (SMILES), Graph representation is more general and intuitive\cite{molecularfingerprints2015}. Graph neural networks (GNNs) provide a promising tool for dealing with molecular graphs and have achieved remarkable success in molecular property prediction\cite{schnet20172}. 
As one of the most widely used GNNs, Message Passing Neural Network (MPNN) operates by aggregating features between adjacent nodes\cite{MPNN2017} and provides a common framework for mainstream GNNs, such as Graph Convolutional Network (GCN)\cite{defferrard2016convolutional} and Graph Attention Network (GAT)\cite{GAT2018}.
More and more chemists have applied this technology to build end-to-end models for predicting the properties of molecules and designing new materials.
However, it has been observed that its topological expressive power is limited by the Weisfeiler-Lehman (WL) isomorphism test\cite{xu2018how}. 
As shown in \cref{fig1}, there exist two nonisomorphic graphs that have different topological structures, and the WL test generates identical coloring and fails to distinguish them.
This reveals MPNN also encounters such insurmountable limitations and affects the performance on the molecular graph.

To solve this problem, in this paper, we extend the message-passing paradigm to a higher-order simplicial complex and propose a Simplicial Message Passing (SMP) to better capture topological information from molecular graphs. The main contributions of this paper are as below:

\begin{itemize}
    \item We propose an SMP framework for aggregating the information from arbitrary order simplex, which can capture the complex interaction and intrinsic topological structure from low-order simplex to high-order simplex.
    \item We then develop a deep learning model following the SMP framework, called Simplicial Message Passing Neural Network (SMPNN), for quantum chemical properties prediction on both topological structure data and geometric data.
    \item We perform a series of experiments on quantum chemistry, including molecular properties prediction and molecular dynamics, and achieve state-of-the-art results.
\end{itemize}
In summary, involving message passing between higher-order simplices is crucial for capturing topological information hidden in complex systems. 
It provides insights into structural connectivity, chemical reactivity, electronic behavior, stability, symmetry, and property-structure relationships. Incorporating topological considerations allows for more accurate and comprehensive predictions, aiding in the discovery and design of materials with tailored properties for various applications. We believe this work not only equips chemists with a powerful tool for discovering new materials, but also provides data scientists and AI developers with a promising perspective and direction.

\section{Related Work}

\subsection{Deep Learning on Chemical Property
Prediction}
Machine learning algorithms are widely used to construct empirical potential and make increased efforts to improve computing accuracy and generalization capacity\cite{overview2,jiangoverview,PTSD2020}. Among numerous machine learning models, neural networks (NN) become the first choice for Potential Energy Surface (PES) calculation due to their powerful nonlinear fitting ability. Most research works focused on designing an effective molecular descriptor for the input of NN. Behler and Parrinello proposed a generalized neural network framework in which a symmetric function was designed to handle the input with different sizes\cite{BPNN2007}, and this method achieved \textit{ab initio} accuracy while it was several orders of magnitude faster. With further exploration of invariance regarding the particular molecule permutation group,  Guo et al. proposed PIP-NN method\cite{PIP12013} which used the Permutation Invariant Polynomial (PIP)\cite{PIP32010} as the input of the neural network and accurately reproduced the analytical potential energy surfaces for $\rm H + H_2$ and $\rm Cl + H_2$ systems. To reduce the number of polynomial inputs in PIP-NN, Fundamental Invariant Neural Network (FI-NN)\cite{FINN2020,FINN2022} introduced an efficient representation of molecular permutation symmetry and broadened the application range for larger molecular systems. Deep Potential (DP)\cite{DP2017}, a simple deep neural network representation of PES, eliminated ad hoc approximation of the above work by taking each atom into a single sub-network while respecting the symmetry of the system. Embedded Atom Neural Network (EANN)\cite{EANN2019} replaced the scalar embedded atom density in empirical Embedded Atom Method (EAM)\cite{EAM1986} with a Gaussian-type orbital based density vector which was sent into NN to obtain atomic energy. Gaussian Moment Neural Network\cite{GMNN2020} used an extendable invariant local molecular descriptor constructed from geometric moments as input of NN to get a high accuracy and efficiency model.   

From the perspective of AI developers and data scientists, this hand-crafted input of NN is less intuitive and violates the end-to-end paradigm of deep learning. Therefore, Graph is adopted to encode the molecules and Graph Neural Network (GNN) is used as the backbone to follow the data-driven manner\cite{PGNN2019, ChemiNet2019, JTVAE2018, GraphDTA2021}. Deep Tensor Neural Network (DTNN)\cite{dtnn2017} treated a molecule as a graph, and the molecular structure was encoded with an inter-atomic distance matrix and an atomic number vector so that the model could capture the interaction between atoms in a pairwise manner. A general framework named Message Passing Neural Networks (MPNN)\cite{MPNN2017} was distilled by reformulating existing models to exploit extra variations. Substantial refinement had been made on the basis of MPNN to improve the performance while reducing the model size and inference time, such as Polarizable Atom Interaction Neural Network (PAINN)\cite{painn2021}
and SpookyNet\cite{spookynet2021}. On the other hand, following the success of the attention mechanism in Natural Language Processing (NLP) and Computer Vision (CV), Graph Attention Network\cite{GAT2018} introduced the attention mechanism into GNN, which made it possible to distinguish the importance of neighbors and its own nodes during feature aggregation, rather than the averaging representation of adjacent nodes in MPNN.
Many works had involved graph attention mechanism for molecular representation and demonstrated that the attention mechanism could effectively extract the nonlocal intramolecular interactions\cite{attn_fp, coley2019graph, CrabNetwang2021, louis2020graph}.

\subsection{Higher-Order Interaction in Complex System}
Simplicial complexes and hypergraphs are natural candidates to equip a mathematical framework for describing group interactions in complex systems\cite{battiston2020networks, kahle2011random}. Early research of random simplicial complexes, Linial–Meshulam model, is simply a higher-order refinement based on the Erdos Renyi (ER) model\cite{linial2006homological}. In this model, the connected graph of $n$ nodes is used as an initialization of m triangles which are formed by three nodes. 
Many variants based on this model focus on the topological data analysis\cite{kahle2014topology}. 
However, with the blossoming of the research on geometric deep learning\cite{bronstein2017geometric}, the simplicial complex provides a novel theoretical perspective for GNNs.
Motivated by Hodge-de Rham theory, Stefania Ebli et al. defined a simplicial convolutional operation to generalize the GNNs to process data living on simplicial complexes\cite{snn}. Corman et al. proposed the element-specific persistent homology (ESPH) method to represent 3D complex geometry by one-dimensional topological invariants\cite{cang2017topologynet}. The combination of ESPH and deep convolutional neural networks exhibited the favorable potential for retaining important biological information. Moreover, to improve the expressive power of GNNs, which is equivalent to the WL graph isomorphism test. 
Message-passing frameworks on simplicial complexes and cell complexes were proposed to capture the multi-level interactions presented in many complex systems\cite{bodnar2021mpsn,bodnar2021cwn}. 
Specifically, the authors of \cite{bodnar2021mpsn} redefined four types of adjacent simplices to expand the perception of local structure and enhance expressive power. Different from these approaches, we just generalize the adjacent definition to higher-order simplex without increasing any adjacent type. Therefore, the vanilla message passing framework of the graph can be treated as a special case in our generalization.

\section{Model Architecture}
In this section, we present the SMP framework and the architecture of SMPNN. The later is 
%
referred to as a deep learning model in the SMP framework.
Firstly, we give a brief introduction to simplicial complex.
Then, we review the message passing framework and describe the main idea of generalizing the concept to arbitrary order simlicial complex.
Finally, we describe the SMPNN architecture in detail.
The overall structure of the SMPNN is illustrated in \cref{fig3}.

\subsection{Preliminary to Simplicial Complex}
In this section, we briefly introduce the concept of simplicial complex, which is a set constructed by piecing simplices together. In discrete geometry, simplex is a fundamental concept used to generalize the notion of a triangle or tetrahedron to arbitrary dimensions. For example, the node of a graph can be treated as a $0$-simplex, and the edge of a graph can be treated as a $1$-simplex. 

Moreover, for a $k$-simplex $\sigma^k$, the boundary $\partial \sigma^k$ of $\sigma^k$ is the closure of the set of all simplices $\sigma^{k-1}$. 
For example, the boundary of a 2-simplex (triangle) $\{v_1,v_2,v_3\}$ is the set that contains all three 1-simplex (edge) $\{v_1,v_2\}$, $\{v_2,v_3\}$, $\{v_3,v_1\}$ of this triangle, and the boundary of a 1-simplex (edge) is the set of two 0-simplex (vertex). 

\begin{definition} If a $k$-simplex $\sigma^k$ is the boundary of a $(k+1)$-simplex $\tau^{k+1}$, we say $\sigma^k \prec \tau^{k+1}$.
\label{def1}
\end{definition}

\subsection{Message Passing Framework for k-simplex}
\label{sec:3.2}
A common undirected graph $\mathcal{G} = \left\{\mathcal{V}, \mathcal{E}\right\}$ contains the node features $h_v$ and edge features $e_vw$. 
The message passing function $M_t$ and the update function $U_t$ consist of a complete $t$-th message passing phase which totally runs for $T$ time steps. The features $h^{t}_v$ of node $v$ in the $t$-th iteration can be written as:

\begin{equation} \label{eq1}
\begin{split}
&m^{t}_v= \sum_{w\in N(v)} M_{t-1}(h^{t-1}_v,h^{t-1}_w,e_{vw}) \\
&h^{t}_v= U_t(h^{t-1}_v, m^{t}_v)  
\end{split}
\end{equation}

\begin{figure}[h]
  \centering
  \includegraphics[width=\linewidth]{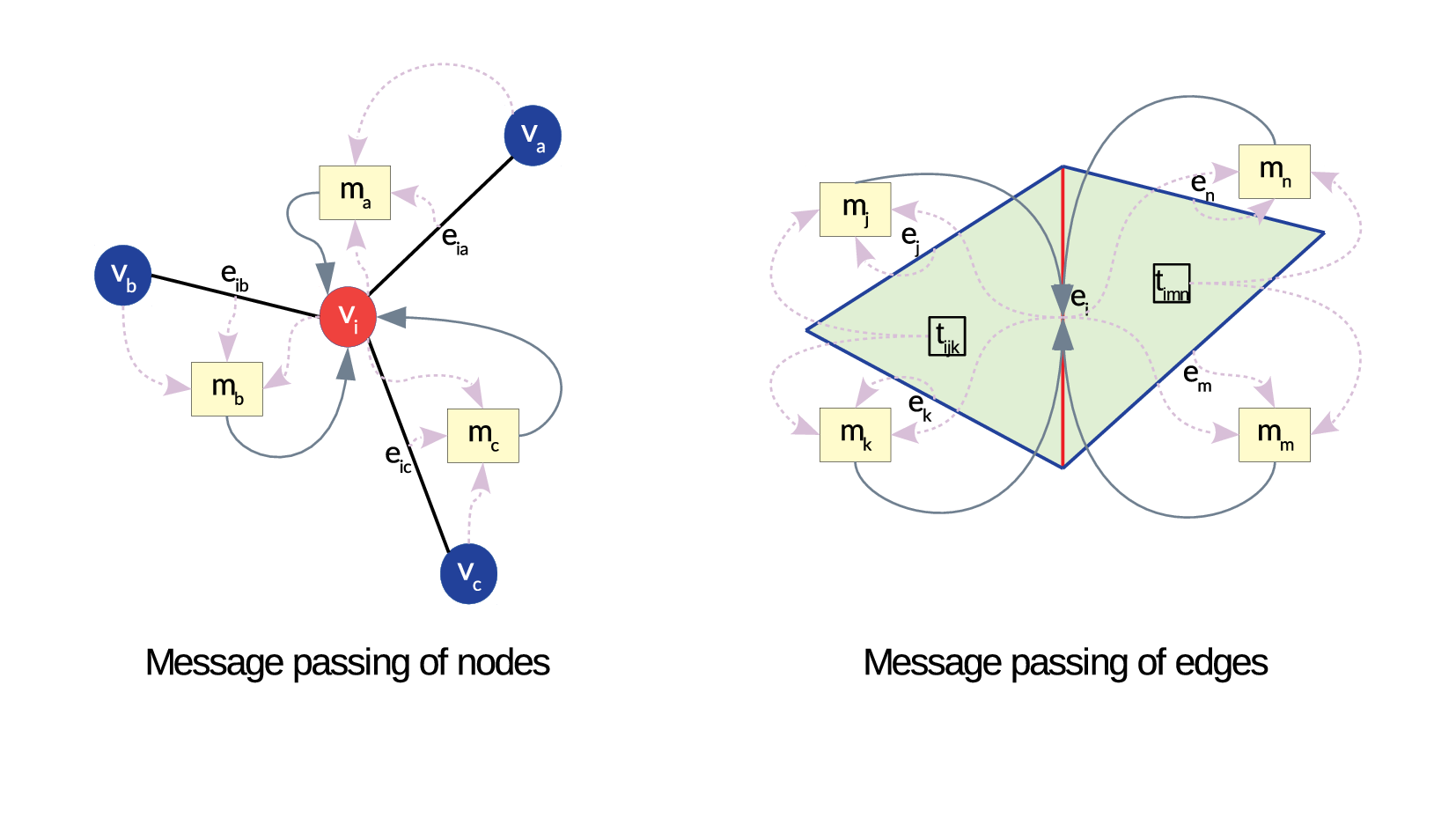}
  \caption{Message passing paradigm for simplices. The left demonstrates the message passing between nodes, and the right demonstrates the message passing between edges.} 
 \label{fig2}
\end{figure}

We treat the graph as a simplicial complex and denote the node $\mathcal{V}$ as 0-simplex $\sigma^0$ and edge $\mathcal{E}$ as 1-simplex $\sigma^1$. Similar to how messages are passed along the edges of a graph, the message-passing mechanism in a higher-order simplicial complex involves passing messages along the "edges" between two simplices. For example, as shown in \cref{fig2}, the message passing between two 0-simplices is along a 1-simplex. 
The message passing between three 1-simplices is along the 2-simplices, that is the triangle constructed by them. Therefore, we generalize the concept of message passing between simplices as \cref{def1}.

\begin{definition} 
If a $k$-simplex $\sigma^k_i \prec \tau^{k+1}$, then message passing to $\sigma^k_i$ is along $\tau^{k+1}$ and involves all the $k$-simplex $\sigma^k \prec \tau^{k+1}$.
\label{def2}
\end{definition}

According to \cref{def2}, we can generalize the vanilla message passing framework to SMP and rewrite the equation \ref{eq1} as:

\begin{equation} \label{eq2}
\begin{split}
&m^{t}_{\sigma^k}= \sum_{\sigma^k_i \prec \tau^{k+1}} M_{t-1}(h^{t-1}(\sigma^k_i),h^{t-1}(\tau^{k+1}))\\
&h^{t}_{\sigma^k}= U_t(h^{t-1}_{\sigma^k}, m^{t}_{\sigma^k})
\end{split}
\end{equation}

\subsection{Simplicial Message Passing Neural Network}
According to the proposed SMP framework in \cref{sec:3.2}, we elaborate SMP to process the downstream tasks of graphs, such as graph classification, node classification, and graph reconstruction. 
Instead of just passing messages between neighboring nodes like MPNN, we need to pass messages between neighboring simplices of different orders, referred to as an SMPNN network in the following. 
The whole SMPNN network can be divided into three steps. 

Firstly, according to the specific tasks. Each simplex in the complex is assigned an initial representation or embedding. For example, in a molecular property prediction task, the node feature can be the atomic number, the edge feature can be the bonding length, the mesh feature can be the area or normal vector of the triangle, and the solid feature can be the volume of the tetrahedron.

Secondly, for each simplex, messages are computed based on the representations of its neighboring simplices. The message computation step involves applying a learnable function or neural network layer that combines the representations of the neighboring simplices in a meaningful way.
Then the computed messages from neighboring simplices are aggregated for each simplex. 
The aggregation process combines the received messages to form a summary representation that captures information from the neighborhood. 
The aggregated messages are then used to update the representation of each simplex. 
This update function takes into account the simplex's current representation and the aggregated messages to generate an updated representation. 
The updated function can be a simple operation like concatenation or a more complex neural network layer.
This step is repeated iteratively for a fixed number of steps or until convergence. 
This mechanism enables us to incorporate knowledge from the local and global neighborhoods of higher-order simplices and can be used for a variety of applications, such as shape analysis, topological data analysis, and protein structure prediction. The application in this paper refers specifically to the molecular property prediction task.

Finally, after iteratively passing and updating messages between simplices, we can capture and propagate information throughout the higher-order simplicial complex. An output layer is elaborated for different tasks. The details of the output layer are discussed in \cref{sec:output_layer}.

As shown in \cref{fig3}, we input simplicial complex into the proposed model instead of a graph. 
Since the molecule has strong 3D structures,
we constrain the order of simplex to $2$. 
The features of simplex will be embedded into high dimensions by an MLP layer to enhance the expressivity. 
Then, the embedded features are sent into a message passing block. Notably, the $k$-simplicial message passing requires the features of $k+1$ simplices, which forge a hierarchical structure. After several iterations, for prediction tasks of topological structure data, the features of $k$-simplices $x^k \in \mathbb{R}^{n\times d}$ will be sent into the pooling layer and concat layer to obtain the final representations $x^k \in \mathbb{R}^{d\times 3}$. And for prediction tasks of geometric data, only the features of 0-simplices are sent into the output block. 


\subsubsection{Output Layer}
\label{sec:output_layer}
\begin{figure}[h]
  \centering
  \includegraphics[width=\linewidth]{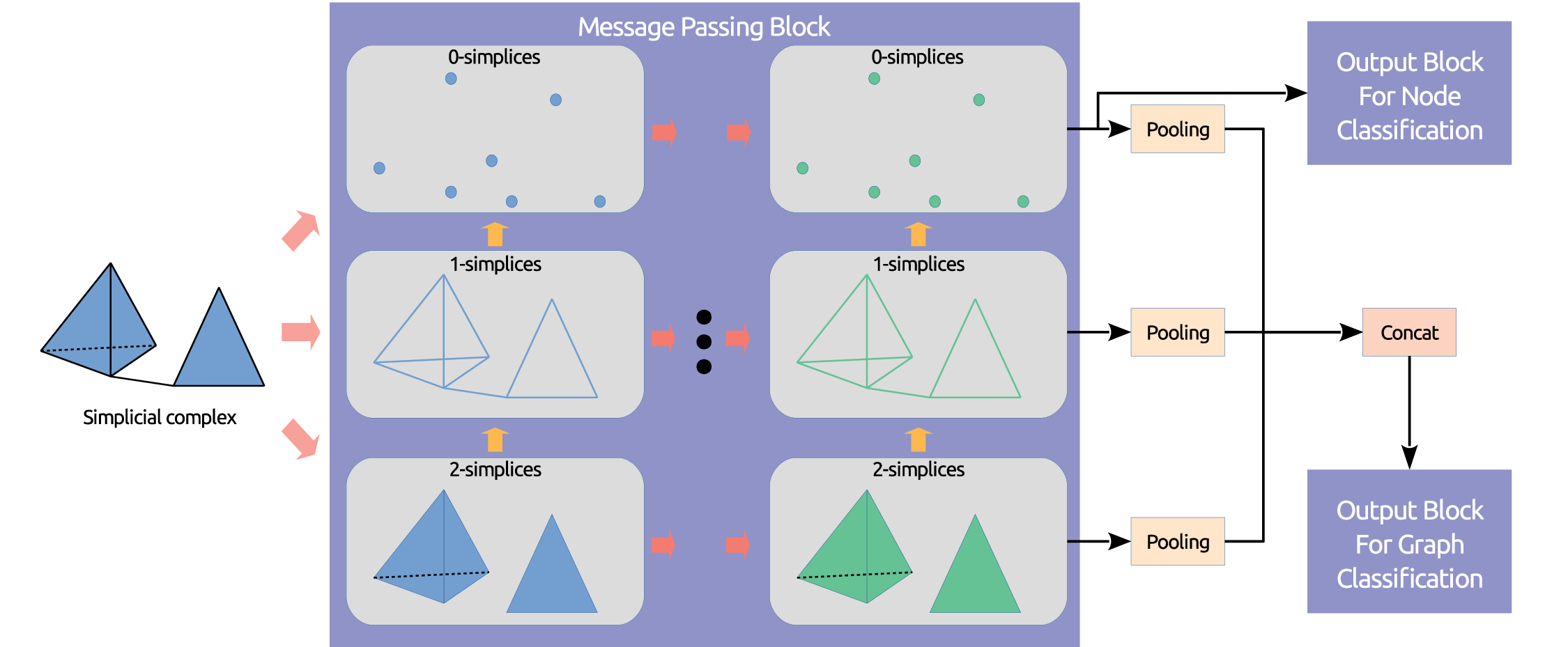}
  \caption{The architecture of proposed SMPNN.} 
  \label{fig3}
\end{figure}

We divide these tasks into two categories, the node classification tasks and the graph classification tasks. And two output blocks are designed for these two different downstream tasks. For graph classification tasks of chemical property prediction, we input the topological structure of molecules, which is usually encoded by SMILES,
and output a scalar result, such as solubility, and drug efficacy. As shown in \cref{fig3}, the output of message passing block, the features of $k$-simplices will be sent into the pooling layer, respectively. Then all the features will be concatenated and sent into an MLP layer for task prediction. 
On the other hand, for node classification tasks of chemical property prediction, the input often contains the Cartesian coordinates of every atom in the molecule. The output results, such as potential energy and force, can be represented as a sum of atomic contributions. Hence, we treat these tasks as node classification/regression tasks. The output block for these tasks is only to receive the features of $0$-simplice and use an MLP layer to calculate the atomic contribution, then sum up them for the total output result.

\section{Experiments}
In this section, we evaluate the performance of SMPNN on several popular benchmarks of chemical properties prediction. 
%
First, we conduct the experiments on topological structure data in \cref{sec4.2}. 
Then, in \cref{sec4.3}, we evaluate the performances of SMPNN  on geometric data, which is also called 3D molecules in \cite{liu2022spherical}, including quantum chemical properties prediction and molecular dynamics. 
Finally, an ablation study of SMPNN is conducted to investigate the contribution of higher-order SMP and the influence of model hyperparameters.

\subsection{Experimental Configuration} 
The hardware environment used in the experiment is Core i7-6700 CPU@3.40 GHz, GeForce GTX1080Ti GPU. The algorithm is developed by Python3.8, and PyTorch1.10 is used as the backend.
We use ADAM\cite{ADAM2015} optimizer for model training with an initial learning rate $1\times10^{-4}$, which is adjusted by cosine decay schedule\cite{SGDR2017}. The mini-batch size is set to 100, and the number of SMP blocks is set to $6$, the embedding dimension is set to 128.

\subsection{Chemical  Property Prediction with Topological Data}
\label{sec4.2}
To verify the effectiveness of SMPNN, we ran several experiments to compare the predictive performance on the solubility of our work and the popular benchmarks. 
These experiments used five public datasets, Delaney\cite{Delaney}, Huuskonen\cite{Huuskonen}, OCHEM\cite{OCHEM}, Tang\cite{tang}, and Cui\cite{cui}. 
All the datasets were using SMILES as input, and we converted the SMILES into graphs in advance. 
The experimental results are shown in \cref{table_sol}.
The results show that our model achieves state-of-the-art performance in all the tasks. 
This confirms the better capacity of our model on capturing the topological structure.

\begin{table}[h]
  \centering
  \caption{Mean predictive accuracy of solubility on five public datasets. The unit used in the experiment is log Mol$/$L.}
    \label{table_sol}
  \begin{tabular}{lcccccc}
    \hline
Dataset & Num of Sample & GCN & GAT & MPNN & D-MPNN & SMPNN  \\
\hline
Delaney\cite{Delaney}     & 1144 &  0.878 & 1.116 & 0.903 & 0.725 & \textbf{0.709} \\ 
Huuskonen\cite{Huuskonen}   & 1297 &  0.917 & 0.967 & 0.723 & 0.714 & \textbf{0.687} \\ 
OCHEM\cite{OCHEM}       & 1311 &  0.952 & 0.788 & 0.675 & 0.693 & \textbf{0.508} \\
Tang\cite{tang}        & 4200 &  0.719 & 1.027 & 0.704 & 0.669 & \textbf{0.566} \\ 
Cui\cite{cui}         & 9943 &  1.072 & 1.017 & 0.983 & 1.023 & \textbf{0.957} \\ 
\hline
  \end{tabular}
\end{table}

\subsection{Chemical  Property Prediction with Geometric Data} 
\label{sec4.3}
There is another type of molecular data in chemical properties prediction, which provides the 3D Cartesian coordinates of every atom in the molecule. Some researchers call it 3D molecular graph while we use the term "Geometric data" to distinguish it from chemical data that only have topological structure.
In order to verify the effectiveness of the proposed algorithm for geometric data in this paper, we chose two public datasets, MD17\cite{MD17} and QM9\cite{QM9,GDB17} to conduct experiments. The molecular dynamics (MD17) datasets provide both the energy and atomic forces of eight small organic molecules as well as the corresponding atom coordinates of the thermalized system. The datasets range in size from 150k to nearly 1M conformational geometries. All trajectories are calculated at a temperature of 500 K and a resolution of $0.5$ fs. The ground truth data are computed via molecular dynamics simulations using PBE+vdW-TS electronic structure method. In our experiment, all eight kinds of organic molecules were trained with the goal of providing highly accurate energy and force predictions. Another dataset QM9 contains $133,885$ stable small organic molecules with up to $7$ heavy atoms (C, O, N, F) and up to 29 atoms in total including H. The ground truth is $12$ quantum chemistry target properties, which are calculated by \textit{ab initio} quantum chemistry methods at the $B3LYP/6-31G(2df, p)$ level. In the experiment of both datasets, the cutoff radius is set to 5\AA.

\begin{table*}[ht]
	\centering
	\caption{Mean absolute error per molecule of predictions for different target properties of the QM9 dataset using 110k training examples. The lowest error is emphasized in bold. 
 The results from Provably powerful graph networks (PPGN)\cite{PGNN2019}, SchNet\cite{schnet20171} and enn-s2s\cite{MPNN2017} are compared.}
	\label{qm9}
	\scalebox{1}{
		\begin{tabular}{l@{\extracolsep{0.2cm}}lcccccc} \ \\
			\hline
		& Target  & Unit & PPGN & SchNet & enn-s2s & SMP     \\ \hline
		& $\epsilon_{homo} $   & $meV$  & 40 & 41 & 43 & $\mathbf{33}$ 
   \\ 
            & $\epsilon_{lumo} $  & $meV$ & 33 & 34 & 37 & $\mathbf{27}$ 
  \\ 
            & $\Delta\epsilon $   & $meV$ & 60 & 63 & 69 & $\mathbf{54}$   \\ 
            & $ZPVE$    & $meV$ & 3.12 & 1.7 &$\mathbf{1.5}$  & 1.6 \\ 
            & $\mu$     & $Debye$  & 0.047 & 0.033 &$\mathbf{0.030}$  & 0.042 \\
            & $\alpha$     & $Bohr^3$ & 0.131 & 0.235 & 0.092 & $\mathbf{0.086}$  \\ 
            & $<R2>$   & $Bohr^2$ & 0.592 &$\mathbf{0.073}$  & 0.180 & 0.241 \\ 
            & $U_0 $   & $meV$   & 37 & 14 & 19 & $\mathbf{12}$ \\ 
            & $U$   & $meV$  & 37 & 19 & 19 & $\mathbf{15}$  \\
	      & $H$    & $meV$  & 36 & $\mathbf{14}$ & 17 & $\mathbf{14}$ \\ 
            & $G$     & $meV$  & 36 & $\mathbf{14}$ & 19 & $\mathbf{14}$ \\
            & $C_v$   & $cal/molK$ & 0.055 & 0.033 & 0.040 & $\mathbf{0.030}$  \\
		\hline
		\end{tabular}
	}
\end{table*}

\begin{table*}[bht]
	\centering
	\caption{Mean Absolute Errors for energy and force prediction in kcal/mol and kcal/mol/\AA\, on MD17 dataset, with $N=1000$. The results provided by GM-sNN\cite{GMNN2020}, EANN\cite{EANN2019}, SchNet\cite{schnet20171} and PhysNet\cite{PhysNet2019} are compared. EANN does not provide results on Benzene molecule.}
	\label{table_1}
	\scalebox{1}{
		\begin{tabular}{p{2cm}@{\extracolsep{0.3cm}}ccccccccc} \ \\
			\hline 
     &  & \multicolumn{4}{c}{N=1000}   \\ \cline{3-6} 
  &  & Schnet & EANN & GM-sNN & SMPNN   \\
	\hline
    \multirow{2}{4em}{Benzene} & Energy &0.08  & - &0.08  &$\mathbf{0.06}$ 
  \\
	&Force &0.31  & - &$\mathbf{0.21}$  & 0.34 \\
 
    \cline{1-2} \cline{3-6} 
    \multirow{2}{4em}{Toluene} & Energy &0.12  & $\mathbf{0.11}$ &0.15 &$\mathbf{0.11}$  \\
	& Force &0.57  & 0.38 &$\mathbf{0.34}$  & 0.42  \\
 
    \cline{1-2} \cline{3-6} 
    \multirow{2}{4em}{Malonaldehyde} & Energy &0.13  & 0.14 &0.12  & $\mathbf{0.10}$  \\
	& Force &0.66  & 0.62 &0.45  & $\mathbf{0.40}$   \\
       
    \cline{1-2} \cline{3-6} 
    \multirow{2}{4em}{Salicylic acid} & Energy &0.20  & $\mathbf{0.14}$ &0.19  & 0.16 \\
	& Force &0.85  & 0.51 &$\mathbf{0.49}$  & 0.52
         \\

    \cline{1-2} \cline{3-6} 
    \multirow{2}{4em}{Aspirin} & Energy &0.37  &  0.33 &0.38  & $\mathbf{0.32}$  \\
	& Force &1.35  & 0.99 &$\mathbf{0.69}$  & 0.99  \\

    \cline{1-2} \cline{3-6} 
    \multirow{2}{4em}{Ethanol} & Energy &0.08  & 0.10 &0.10  & $\mathbf{0.07}$ \\
	& Force &0.39  & 0.47 &0.33  &  $\mathbf{0.22}$  \\

    \cline{1-2} \cline{3-6} 
    \multirow{2}{4em}{Uracil} & Energy &0.14  & $\mathbf{0.11}$  &0.12  & $\mathbf{0.11}$   \\
	& Force & 0.56  & 0.35 &$\mathbf{0.33}$   & 0.36  \\

    \cline{1-2} \cline{3-6} 
    \multirow{2}{4em}{Naphthalene} & Energy &0.16  & $\mathbf{0.12}$ &0.17  & 0.15  \\
	& Force &0.58  & $\mathbf{0.27}$ &0.36  & 0.36  \\   
		\hline 
		\end{tabular}
	}
\end{table*}
\begin{table*}[bht]
	\centering
	\caption{Mean Absolute Errors for energy and force prediction in kcal/mol and kcal/mol/\AA\, on MD17 dataset, with $N=50000$. The results provided by GM-sNN\cite{GMNN2020}, EANN\cite{EANN2019}, SchNet\cite{schnet20171} and PhysNet\cite{PhysNet2019} are compared. EANN does not provide results on Benzene molecule.}
	\label{table_12}
	\scalebox{1}{
		\begin{tabular}{p{2cm}@{\extracolsep{0.3cm}}ccccccccc} \ \\
			\hline 
     &  & \multicolumn{4}{c}{N=50000}  \\ \cline{3-6} 
  &   & Schnet & PhysNet & GM-sNN & SMP  \\
	\hline
    \multirow{2}{4em}{Benzene} & Energy 
 &$\mathbf{0.07}$  &  $\mathbf{0.07}$ & $\mathbf{0.07}$  & 0.10 \\
	&Force &0.17  & 0.15 &$\mathbf{0.14}$  & 0.17 \\
 
    \cline{1-2} \cline{3-6} 
    \multirow{2}{4em}{Toluene} & Energy  & 0.09  & 0.10   & 0.14  & $\mathbf{0.07}$ \\
	& Force  &0.09  & $\mathbf{0.03}$ &0.10  & 0.05 \\
 
    \cline{1-2} \cline{3-6} 
    \multirow{2}{4em}{Malonaldehyde} & Energy &0.08  & $\mathbf{0.07}$ &0.12  & $\mathbf{0.07}$ \\
	& Force  & 0.08  & $\mathbf{0.04}$ &0.08  & $\mathbf{0.04}$ \\
       
    \cline{1-2} \cline{3-6} 
    \multirow{2}{4em}{Salicylic acid} & Energy  &0.10 & 0.11 &0.19  &$\mathbf{ 0.09}$ \\
	& Force 
        & 0.19  & $\mathbf{0.04}$ &0.14  & 0.09 \\

    \cline{1-2} \cline{3-6} 
    \multirow{2}{4em}{Aspirin} & Energy  &0.12  &  0.12 &0.19  & $\mathbf{0.11}$ \\
	& Force  & 0.33  & $\mathbf{0.06}$ & 0.26  & 0.14 \\

    \cline{1-2} \cline{3-6} 
    \multirow{2}{4em}{Ethanol} & Energy  &0.05  & 0.05 &0.05  & $\mathbf{0.04}$ \\
	& Force  &0.05  & 0.03 &0.06  &  $\mathbf{0.02}$ \\

    \cline{1-2} \cline{3-6} 
    \multirow{2}{4em}{Uracil} & Energy   & $\mathbf{0.10}$  & $\mathbf{0.10}$  &$\mathbf{0.10}$  & $\mathbf{0.10}$ \\
	& Force  & 0.11  &$\mathbf{0.03}$ &0.07  & 0.04 \\

    \cline{1-2} \cline{3-6} 
    \multirow{2}{4em}{Naphthalene} & Energy  &0.11  & 0.12 &0.13  &$\mathbf{0.08}$ \\
	& Force  &0.11  &$\mathbf{0.04}$ &0.13  &0.05 \\   
		\hline 
		\end{tabular}
	}
\end{table*}

We randomly choose 1000 and 50000 molecular configurations as the training set and the remaining data as the test set in the MD17 dataset. Mean Absolute Error (MAE) between predictions and ground truth per molecule is applied as evaluation metrics. Firstly, we compare our model with four benchmark  methods (Schnet\cite{schnet20171}, EANN\cite{EANN2019}, GMNN\cite{GMNN2020}, Physnet\cite{PhysNet2019}). The SMPNN model was trained on $N=1000$ and $N=50000$ samples. The results of experiments, as shown in \cref{table_1} and \cref{table_12}, demonstrate SMPNN performs best or at least equal to other models on 9 out of 16 targets with 50000 training samples and on half targets with 1000 training samples. Notably, for energy calculation, SMPNN achieves state-of-the-art performance on all the organic molecules except Benzene with 50000 training samples. 
When the number of training samples is decreased to 1000, SMPNN still outperforms others on most molecules datasets, which fully demonstrates the effectiveness of SMPNN in the prediction of chemical properties on geometric data.

\subsection{Ablation Study } 
To further investigate the influence of higher-order simplex on the expressive power of our model, we removed the $2$-simplicial and $1$-simplicial message-passing block and show the results in \cref{table_3}. 
We carried out ablation experiments and selected two organic compounds with the largest and smallest number of samples from MD17 dataset, Malonaldehyde with $993,237$  data samples and Uracil with $133,770$ data samples. 
We also carried out the experiments on the whole MD17 dataset and QM9 dataset, due to the page limit, the results can be found in the appendix.

As shown in \cref{table_3}, compared to $0$-SMP, which is identical to the traditional MPNN, adding $1$-SMP can substantially enhance the performance on force prediction but provides a trivial improvement on energy prediction, while involving $2$-SMP can remarkably boost the accuracy on both tasks. 
The experimental results indicated that ignoring higher-order interactions, such as the edge and mesh, will restrain the performance of the model. Therefore, we speculated that the intrinsic relationship captured by our model is more proximate to the first principles.

\begin{figure}[h]
  \centering
  \includegraphics[width = \linewidth]{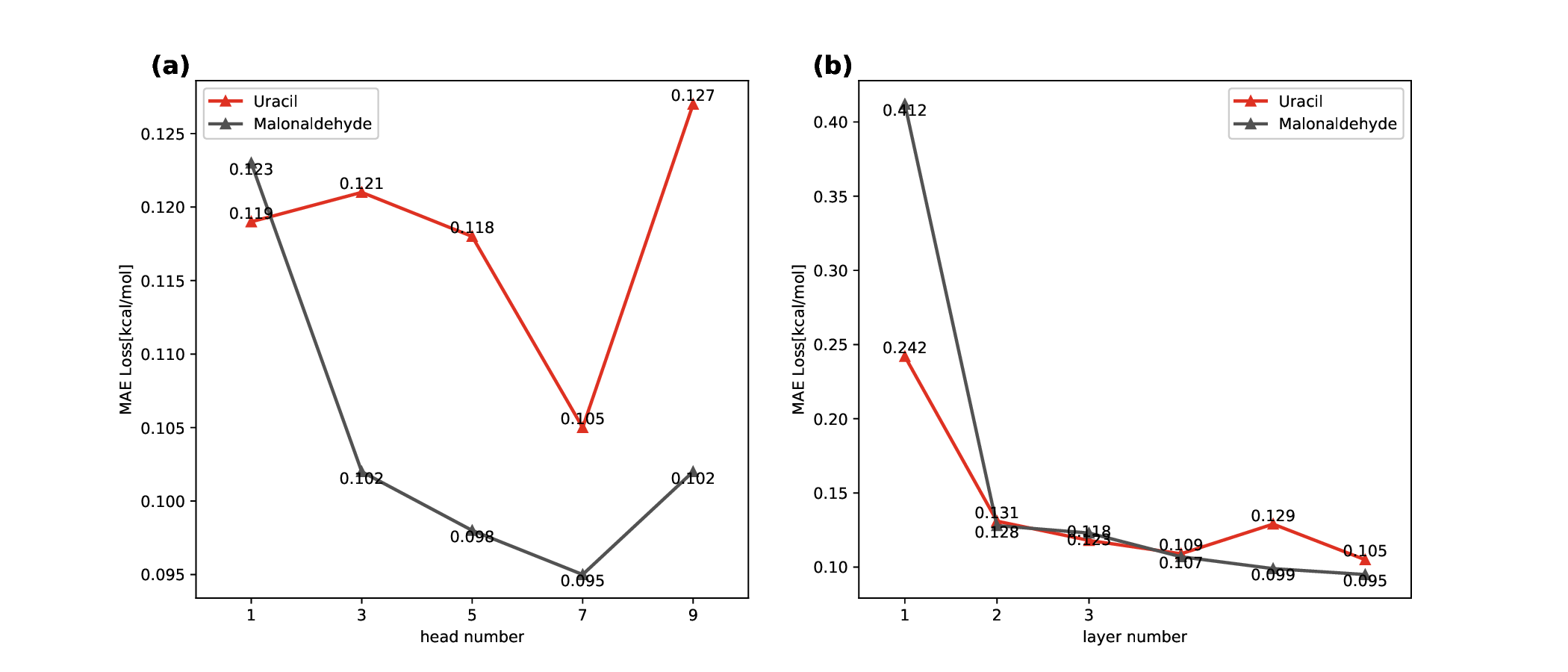}
  \caption{The comparison with width(left) and depth(right) of message-passing block.} 
  \label{fig4}

\end{figure}

Furthermore, we varied the dimension of the embedding features and the number of layers in the message-passing block to investigate the influence of the width and depth of the model. 
As shown in \cref{fig4}(a), we enlarged the model width by increasing the dimension of the embedding features. For a large dataset (Malonaldehyde), the performance improved and then soon saturated, while there was no substantial improvement for a small dataset (Uracil). 
Meanwhile, \cref{fig4}(b) shows that increasing  the number of message-passing layers can give rise to better performance on both datasets. 
However, a trade-off should be considered because stacking too many layers will consume more computational resources and training time.

\begin{table}[htb]
	\centering
	\caption{Results of removing the higher-order simplex message-passing block. Scores are given by MAE of energy (kcal/mol) and force (kcal/mol/\AA\,) prediction. $0$-SMP denotes $0$-simplicial message passing, and the same for other cases.}
	\label{table_3}	
	\begin{tabular}{c@{\extracolsep{0.4cm}}cccccc} 
	\hline
   \multirow{2}{4em}{$0$-SMP} & \multirow{2}{4em}{$1$-SMP}& \multirow{2}{4em}{$2$-SMP} &\multicolumn{2}{c}{Malonaldehyde} &  \multicolumn{2}{c}{Uracil} \\ \cline{4-5} \cline{6-7}
		&  &  &energy &force &energy &force
\\ \hline
	\checkmark	& \xmark   &  \xmark  & 0.13 & 0.66 & 0.14 & 0.56 \\ 
	\checkmark	& \checkmark & \xmark  &0.14   & 0.49    & 0.12   & 0.44 \\ 
	\checkmark	& \checkmark & \checkmark    & 0.10 & 0.44 & 0.11 & 0.36  \\ 
	
		\hline		
		\end{tabular}	
\end{table}

\section{Conclusion}
In this study, we established a framework for message-passing in simplicial complex and demonstrated that our work can provide a comprehensive representation of the molecular topology, allowing for a more nuanced analysis of its structural features. 
The results of experiments showed that, by involving higher-order simplices, our model encompassed both the fine-grained details and the larger-scale patterns, thereby capturing crucial aspects of molecular connectivity and arrangement.
However, the introduction of high-order simplex often consumes more computing resources, and the success of our work is initially verified by the empirical results of experiments. 
In future work, we develop more efficient methods for optimizing the algorithm and delve into the theoretical analysis.
This will provide new explorations for the development of geometric deep learning in chemistry and other related fields.

\medskip

{
\bibliographystyle{unsrt}
\bibliography{ref}}


\end{document}